\documentclass[10pt,twocolumn,letterpaper]{article}

\usepackage{cvpr}
\usepackage{times}
\usepackage{epsfig}
\usepackage{graphicx}
\usepackage{amsmath}
\usepackage{amssymb}

\usepackage[subrefformat=parens]{subcaption}

\usepackage[font=small]{caption}
\usepackage[font=small]{subcaption}
\usepackage{url}
\usepackage{amsmath,  accents}
\usepackage{bm}

\usepackage[pagebackref=true,breaklinks=true,letterpaper=true,colorlinks,bookmarks=false]{hyperref}

\cvprfinalcopy % *** Uncomment this line for the final submission

\def\cvprPaperID{7311} % *** Enter the CVPR Paper ID here

\ifcvprfinal\fi
\begin{document}

%%%%%%%%% TITLE
\title{Unpaired Image Super-Resolution using Pseudo-Supervision}

\author{Shunta Maeda\\
Navier Inc.\\
{\tt\small shunta@navier.co.jp}
% For a paper whose authors are all at the same institution,
% omit the following lines up until the closing ``}''.
% Additional authors and addresses can be added with ``\and'',
% just like the second author.
% To save space, use either the email address or home page, not both
%\and
%Second Author\\
%Institution2\\
%First line of institution2 address\\
%{\tt\small secondauthor@i2.org}
}

\maketitle
%\thispagestyle{empty}

%%%%%%%%% ABSTRACT
\begin{abstract}
In most studies on learning-based image super-resolution (SR), the paired training dataset is created by downscaling high-resolution (HR) images with a predetermined operation (e.g., bicubic).
However, these methods fail to super-resolve real-world low-resolution (LR) images, for which the degradation process is much more complicated and unknown.
In this paper, we propose an unpaired SR method using a generative adversarial network that does not require a paired/aligned training dataset.
Our network consists of an unpaired kernel/noise correction network and a pseudo-paired SR network.
The correction network removes noise and adjusts the kernel of the inputted LR image; then, the corrected clean LR image is upscaled by the SR network.
In the training phase, the correction network also produces a pseudo-clean LR image from the inputted HR image, and then a mapping from the pseudo-clean LR image to the inputted HR image is learned by the SR network in a paired manner.
Because our SR network is independent of the correction network, well-studied existing network architectures and pixel-wise loss functions can be integrated with the proposed framework.
Experiments on diverse datasets show that the proposed method is superior to existing solutions to the unpaired SR problem.
\end{abstract}

%%%%%%%%% BODY TEXT
\begin{figure}[t]
\centering
\includegraphics[width=0.925\linewidth]{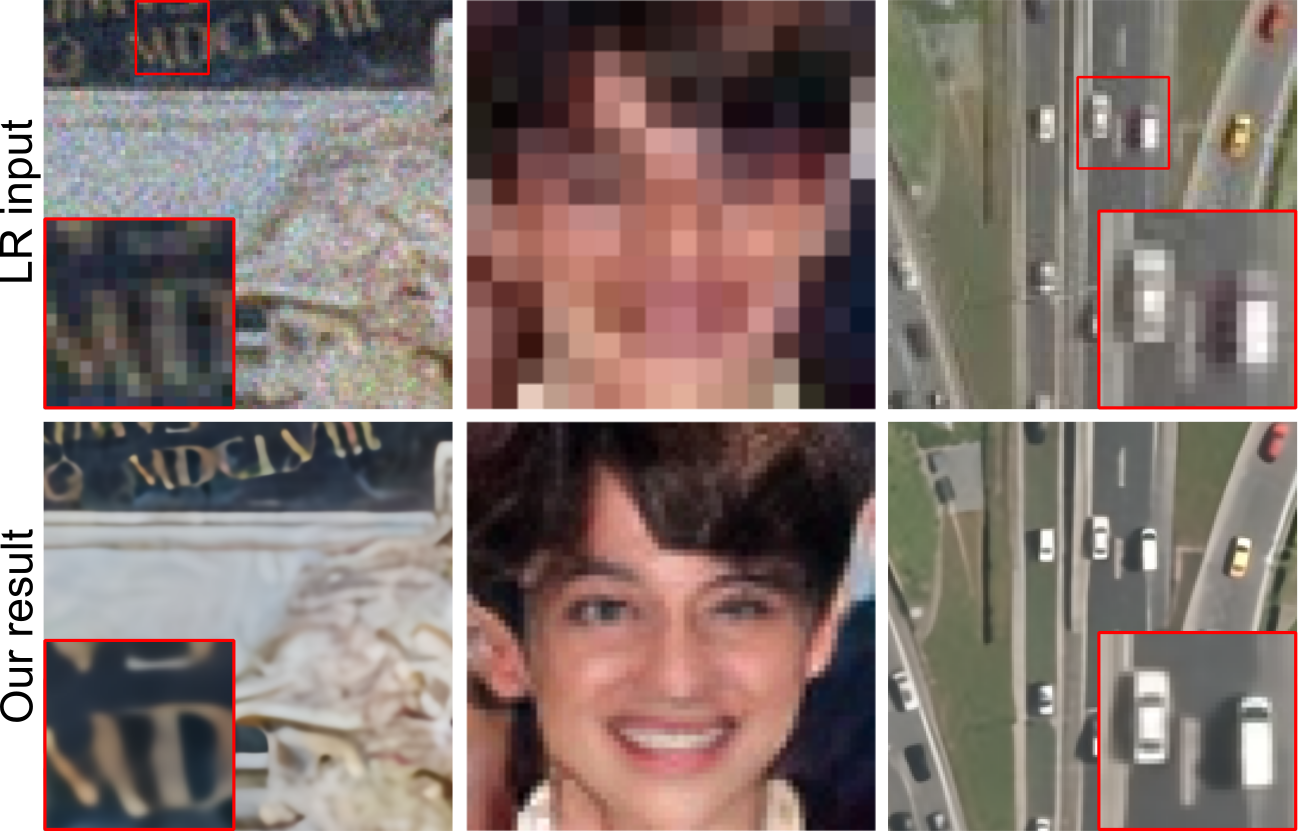}
\caption{{\bf Unpaired SR results on diverse datasets.} From left to right: $\times 4$ SR result for a synthetically degraded LR image from DIV2K realistic-wild set~\cite{timofte2018ntire}, $\times 4$ SR result for a real-world LR face image from Widerface~\cite{yang2016wider}, and $\times 2$ SR result for a real-world LR aerial image from DOTA~\cite{xia2018dota}. Zoom in for better view.}
\label{fig:top}
\end{figure}

\section{Introduction}
Image super-resolution (SR) is a fundamental ill-posed problem in low-level vision that reconstructs a high-resolution (HR) image from its low-resolution (LR) observation.
Recent progress in the on deep learning-based methods has significantly improved the performance of SR, increasing attention from the practical perspective.
However, in many studies, training image pairs are generated by a predetermined downscaling operation (\eg, bicubic) on the HR images.
This method of dataset preparation is not practical in real-world scenarios because there is usually no HR image corresponding to the given LR one.

Some recent studies have proposed methods to overcome the absence of HR--LR image pairs, such as blind SR methods~\cite{shocher2018zero, gu2019blind, zhou2019kernel} and generative adversarial network (GAN)-based unpaired SR methods~\cite{yuan2018unsupervised, bulat2018learn, zhao2018unsupervised, lugmayr2019unsupervised}.
Blind SR aims to reconstruct HR images from LR ones degraded by arbitrary kernels.
Although recent studies have achieved ``blindness'' for limited forms of degradation (\eg, blur), real LR images are not always represented with such degradation; thus, they perform poorly on the images degraded by not expected processes.
By contrast, GAN-based unpaired SR methods can directly learn a mapping from LR to HR images without assuming any degradation processes.

\begin{figure}[t]
\centering
\begin{tabular}{c}
    \begin{minipage}{0.825\hsize}
    \centering
    \includegraphics[width=1.0\linewidth]{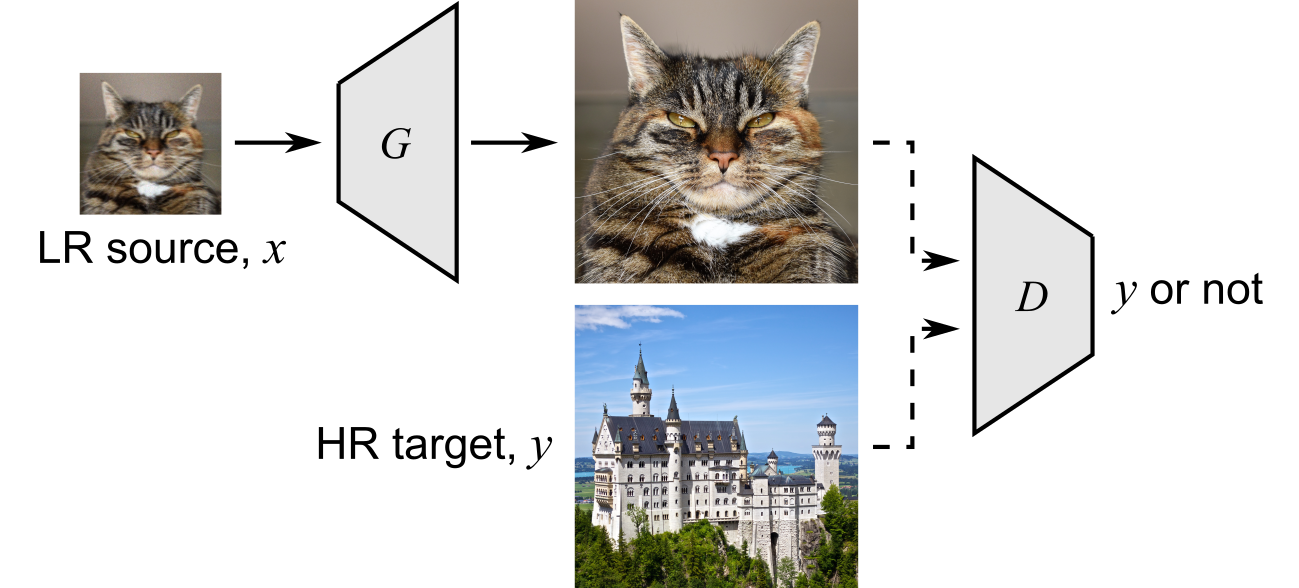}
    \subcaption{Direct approach.}
    \label{fig:two_approaches_a}
    \end{minipage}\vspace{2mm}\\

    \begin{minipage}{0.825\hsize}
    \centering
    \includegraphics[width=1.0\linewidth]{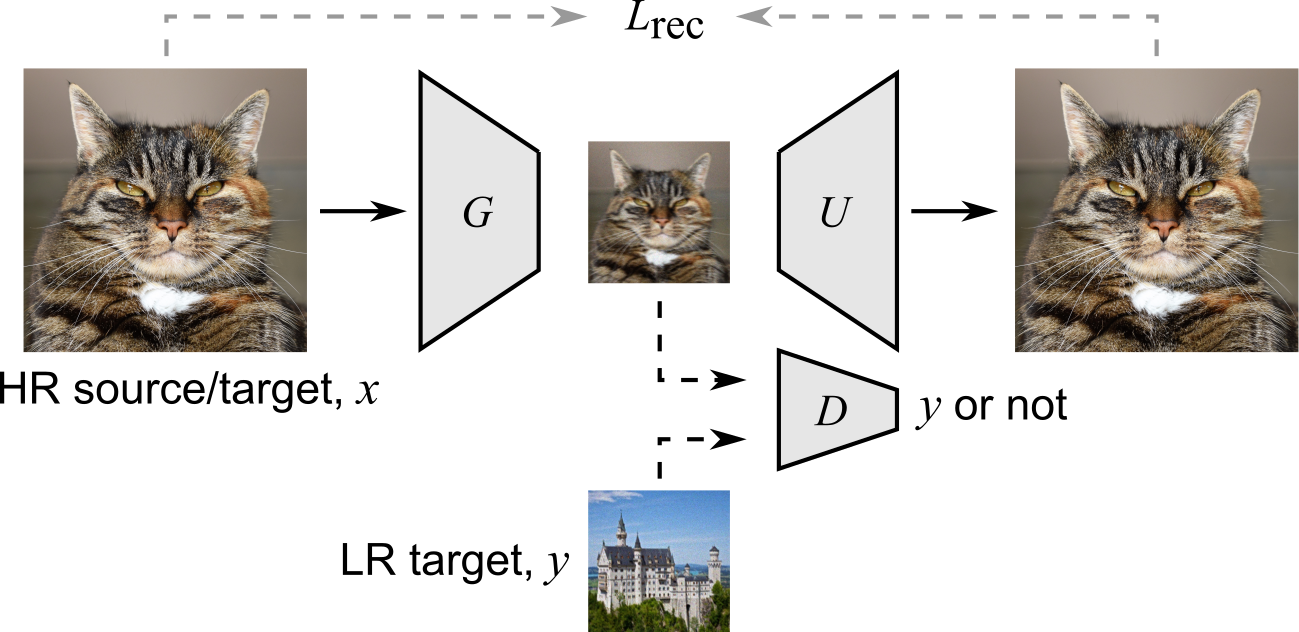}
    \subcaption{Indirect approach.}
    \label{fig:two_approaches_b}
    \end{minipage}
\end{tabular}
\caption{{\bf Two distinct approaches to unpaired SR using GANs.} (a) Generator directly upscales LR images. (b) Generator first downscales HR images and uses the generated LR images to train SR network $U$.}
\label{fig:two_approaches}
\end{figure}

GANs learn to generate images with the same distribution as the target domain through a minimax game between a generator and discriminator~\cite{goodfellow2014generative, salimans2016improved}.
GAN-based unpaired SR methods can be roughly classified according to whether they start from an LR image (direct approach; Fig.~\ref{fig:two_approaches_a}) or an HR image (indirect approach; Fig.~\ref{fig:two_approaches_b}).

\noindent{{\bf Direct approach.\ }}
In this approach, a generator upscales source LR images to fool an HR discriminator~\cite{yuan2018unsupervised}.
The main drawback of this approach is that the pixel-wise loss functions cannot be used to train the generator, \ie, SR network.
In paired SR methods, the pixel-wise loss between reconstructed images and HR target images plays a crucial role not only in distortion-oriented methods but also in perception-oriented methods~\cite{ledig2017photo, blau20182018}.

\noindent{{\bf Indirect approach.\ }}
In this approach, a generator downscales source HR images to fool an LR discriminator~\cite{bulat2018learn, lugmayr2019unsupervised}.
The generated LR images are then used to train the SR network in a paired manner.
The main drawback of this approach is that the deviation between the generated LR distribution and the true LR distribution causes train--test discrepancy, degrading the test time performance.

\begin{figure*}[t]
\centering
\includegraphics[width=0.925\linewidth]{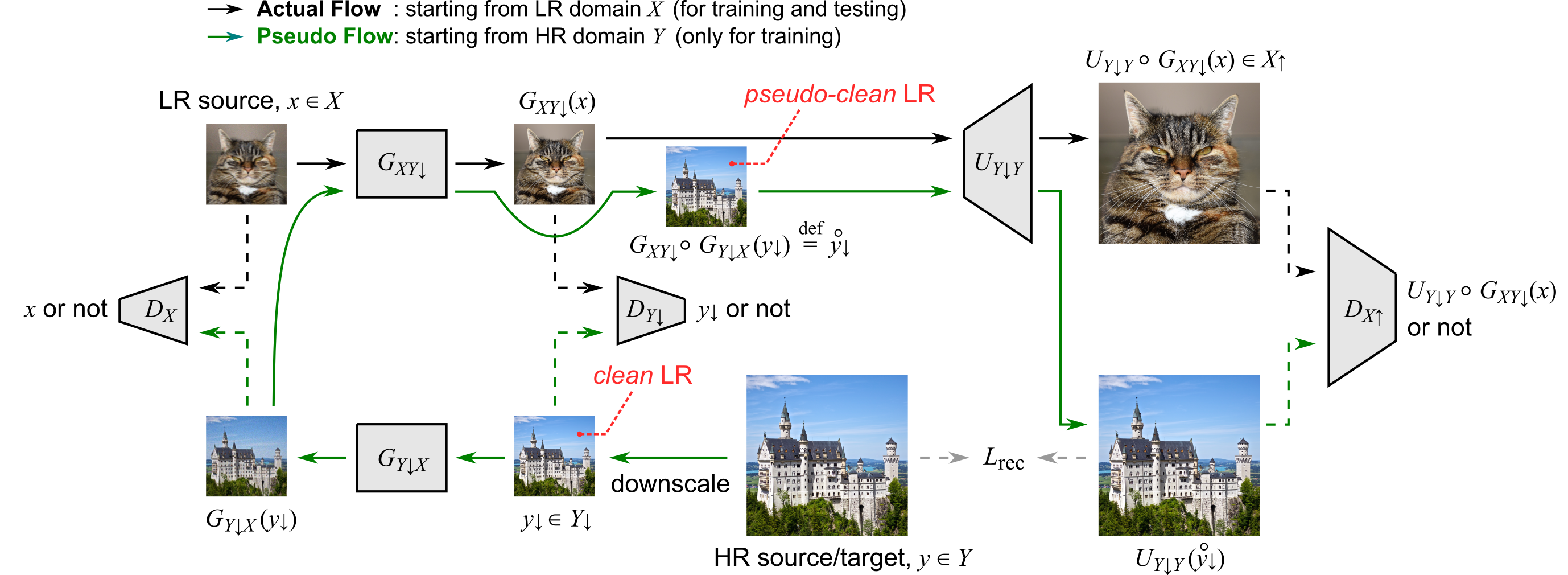}
\caption{{\bf Data-flow diagram of proposed method.} 
SR network $U_{Y_{\downarrow}Y}$ can be learned in a paired manner through $\mathcal{L}_{rec}$, even if the training dataset $\{X, Y\}$ is not paired. The whole network is end-to-end trainable.}
\label{fig:our_approach}
\vspace{-2.5mm}
\end{figure*}

\noindent{{\bf Our approach.\ }}
The main contribution of this work is that we simultaneously overcome the drawbacks of the above two approaches by separating the entire network into an unpaired kernel/noise correction network and a {\it pseudo-paired} SR network (Fig.~\ref{fig:our_approach}).
The correction network is a CycleGAN~\cite{zhu2017unpaired}-based unpaired LR $\leftrightarrow$ {\it clean}~LR translation.
The SR network is a paired {\it clean}~LR $\rightarrow$ HR mapping, where the {\it clean}~LR images are created by downscaling the HR images with a predetermined operation.
In the training phase, the correction network also generates {\it pseudo-clean}~LR images by first mapping the {\it clean}~LR images to the true LR domain and then pulling them back to the {\it clean}~LR domain.
The SR network is learned to reconstruct the original HR images from the {\it pseudo-clean}~LR images in a paired manner.
With the following two merits, our method achieves superior results to state-of-the-arts ones:
(1) Because our correction network is trained on not only the generated LR images but also the true LR images through the bi-directional structure, the deviation between the generated LR distribution and the true LR distribution does not critically degrade the test time performance.
(2) Any existing SR networks and pixel-wise loss functions can be integrated because our SR network is separated to be able to learn in a paired manner.

%------------------------------------------------------------------------

\section{Related Work}
The training data, network architecture, and objective function are three essential elements of a learning deep network.
Paired image SR is aimed at optimizing the network architecture and/or objective function to improve performance under the assumption that ideal training data exists.
However, in many practical cases, there is a lack of training data (\ie, target HR images corresponding to source LR images).
This problem has been addressed by recent studies on blind and unpaired image SR.
As another approach, a few recent works~\cite{chen2019camera, zhang2019zoom, cai2019toward} have built real paired SR datasets using specialized hardware and data correction processes, which are difficult to scale.

\subsection{Paired Image Super-Resolution}
In most SR studies, the paired training dataset is created by downscaling HR images with a predetermined operation (\eg, bicubic).
Since the first convolutional neural network (CNN)-based SR network~\cite{dong2015image}, various SR networks have been proposed to improve LR-to-HR reconstruction performance.
Early studies~\cite{kim2016accurate, lim2017enhanced} found that a deeper network performs better with residual learning.
A proposed residual channel attention network (RCAN)~\cite{zhang2018image} achieved further improved depth and performance.
Upscaling strategies have also been studied, such as progressive upscaling of LapSRN~\cite{lai2017deep} and iterative upscaling and downscaling of DBPN~\cite{haris2018deep}.
In these studies, a simple L1 or L2 distance was used as the objective function, but it is known that these simple distances alone result in blurred textures.
To improve the perceptual quality, SRGAN~\cite{ledig2017photo} introduced perceptual loss~\cite{johnson2016perceptual} and adversarial loss~\cite{goodfellow2014generative}, realizing more visually pleasing results.
ESRGAN~\cite{wang2018esrgan}, which is an enhanced version of SRGAN, is one of the state-of-the-art perception-oriented models.

\subsection{Blind Image Super-Resolution}
Relatively less research attention has been paid to blind image SR despite its importance for practical applications.
Studies on blind SR usually focus on models that are only blind to the blur kernels~\cite{michaeli2013nonparametric, shao2015simple, shocher2018zero, gu2019blind, zhou2019kernel}.
For instance, ZSSR~\cite{shocher2018zero} exploits the recurrence of information inside a single image to upscale images with different blur kernels, and IKC~\cite{gu2019blind} uses the intermediate outputs to iteratively correct the mismatch of blur kernels.
Few studies on blind SR have addressed the combined degradation problem (\ie, additive noise, compression artifacts, etc.) beyond the blur blind SR, whereas several blind methods have been proposed for specific degradation problems, such as denoising~\cite{krull2019noise2void} and motion deblurring~\cite{nah2017deep, kupyn2018deblurgan}.

\subsection{Unpaired Image Super-Resolution}
A few recent works have addressed the SR problem without using a paired training dataset.
Different from the unpaired translation methods, such as CycleGAN~\cite{zhu2017unpaired} and DualGAN~\cite{yi2017dualgan}, unpaired SR aims to upscale source LR images while preserving style and local structure.
Bulat~\etal~\cite{bulat2018learn} and Lugmayr~\etal~\cite{lugmayr2019unsupervised} first trained a high-to-low degradation network and then used the degraded outputs to train a low-to-high SR network.
Yuan~\etal~\cite{yuan2018unsupervised} proposed a cycle-in-cycle network to simultaneously learn a degradation network and an SR network.
Different from our method, the degradation network of Yuan~\etal is deterministic, and the SR network is incorporated with the bi-cycle network; thus, the usable loss function is limited.
Zhao~\etal~\cite{zhao2018unsupervised} also jointly stabilized the training of a degradation network and SR network by utilizing a bi-directional structure.
Similar to Yuan~\etal , the SR network of Zhao~\etal has a limited degree of freedom to select the loss function.

%------------------------------------------------------------------------
\section{Proposed Method}
Our goal is to learn a mapping $F_{XY}$ from an LR source domain $X$ to an HR target domain $Y$ based on the given unpaired training samples $x~(\in X)$ and $y~(\in Y)$.
Here, we define ``{\it clean}~LR,'' \ie, HR images downscaled with a predetermined operation, as $y_{\downarrow}~(\in Y_{\downarrow})$.
The downscaling operation $Y \rightarrow Y_{\downarrow}$ used is a combination of Gaussian blur with $\sigma = ({\rm scale~factor})/2$ and bicubic downscaling.
The mapping $F_{XY}$ of our model is a combination of the two mappings $G_{XY_{\downarrow}}$ and $U_{Y_{\downarrow}Y}$, where $G_{XY_{\downarrow}}$ is a mapping from $X$ to $Y_{\downarrow}$, and $U_{Y_{\downarrow}Y}$ is an upscaling mapping from $Y_{\downarrow}$ to $Y$.
Figure~\ref{fig:our_approach} illustrates the proposed framework.

\noindent{{\bf Domain transfer in LR.\ }}
We use a CycleGAN~\cite{zhu2017unpaired}-based model for the domain transfer in LR. Two generators, $G_{XY_{\downarrow}}$ and its inverse mapping $G_{Y_{\downarrow}X}$, are simultaneously learned to enforce cycle consistency, \ie $G_{XY_{\downarrow}} \circ G_{Y_{\downarrow}X}(y_{\downarrow}) \approx y_{\downarrow}$\footnote{$F \circ G(x) := F(G(x))$.}.
The training of the generator $G_{XY_{\downarrow}}$ ($G_{Y_{\downarrow}X}$) requires a discriminator $D_{Y_{\downarrow}}$ ($D_X$) that aims to detect translated samples from the real examples $y_{\downarrow}$ ($x$).

\noindent{{\bf Mapping from LR to HR.\ }}
The upscaling mapping $U_{Y_{\downarrow}Y}$ is learned to reconstruct HR image $y$ from a {\it pseudo-clean}~LR image $G_{XY_{\downarrow}} \circ G_{Y_{\downarrow}X}(y_{\downarrow})$ in a paired manner.
Thus, any pixel-wise loss functions can be used to train $U_{Y_{\downarrow}Y}$.
Hereafter, we denote $G_{XY_{\downarrow}} \circ G_{Y_{\downarrow}X}(y_{\downarrow})$ as ``$\accentset{\scriptstyle\circ}{y_{\downarrow}}$''.
%Since both of the $x$ and $G_{Y_{\downarrow}X}(y_{\downarrow})$ are independently fed to the $G_{XY_{\downarrow}}$ during the training, some differences of the distributions between $x$ and $G_{Y_{\downarrow}X}(y_{\downarrow})$ do not significantly affect the quality of the test image transformation.

\noindent{{\bf Adjustment with HR Discriminator.\ }}
While $\accentset{\scriptstyle\circ}{y_{\downarrow}}$ is used to train $U_{Y_{\downarrow}Y}$, the actual input at test time is $G_{XY_{\downarrow}}(x)$.
Accordingly, $\accentset{\scriptstyle\circ}{y_{\downarrow}} \sim G_{XY_{\downarrow}}(x)$ is required to minimize the train--test discrepancy.
Although this requirement is satisfied to some extent by the normal CycleGAN, we introduce an additional discriminator $D_{X_{\uparrow}}$, which takes the output of $U_{Y_{\downarrow}Y}$ as input so that $U_{Y_{\downarrow}Y}(\accentset{\scriptstyle\circ}{y_{\downarrow}})$ gets closer to $U_{Y_{\downarrow}Y} \circ G_{XY_{\downarrow}}(x)$.
Here, we define $X_{\uparrow}$ as a domain consisting of $U_{Y_{\downarrow}Y} \circ G_{XY_{\downarrow}}(x)$.
Thus, $X_{\uparrow}$ is an unfixed domain that shifts during training.
Note that $D_{X_{\uparrow}}$ updates the parameters of the two generators, and $U_{Y_{\downarrow}Y}$ is simply used as an amplifier of local image features.

\subsection{Loss Functions}
\noindent{{\bf Adversarial loss.\ }}
We impose an adversarial constraint~\cite{goodfellow2014generative} on both generators $G_{XY_{\downarrow}}$ and $G_{Y_{\downarrow}X}$.
As a specific example, an adversarial loss for $G_{XY_{\downarrow}}$ and $D_{Y_{\downarrow}}$ is expressed as
\begin{equation}
\begin{split}
\mathcal{L}_{adv}(G_{XY_{\downarrow}}, D_{Y_{\downarrow}}, X, Y_{\downarrow}) = \mathbb{E}_{y_{\downarrow} \sim {\rm P}_{Y_{\downarrow}}}[\log D_{Y_{\downarrow}}(y_{\downarrow})]~\\
+~\mathbb{E}_{x \sim {\rm P}_X}[\log (1 - D_{Y_{\downarrow}}(G_{XY_{\downarrow}}(x)))],
\end{split}
\label{eq:loss_adv}
\end{equation}
where ${\rm P}_{X}$ (${\rm P}_{Y_{\downarrow}}$) is the data distribution of the domain $X$ ($Y_{\downarrow}$).
$G_{XY_{\downarrow}}$ and $D_{Y_{\downarrow}}$ simultaneously optimized each other through a mini-max game between them, \ie, $\min_{G_{XY_{\downarrow}}} \max_{D_{Y_{\downarrow}}} \mathcal{L}_{adv}(G_{XY_{\downarrow}}, D_{Y_{\downarrow}}, X, Y_{\downarrow})$.
Similar to the CycleGAN framework, the inverse mapping $G_{Y_{\downarrow}X}$ and its corresponding discriminator $D_X$ are also optimized: $\min_{G_{Y_{\downarrow}X}} \max_{D_{X}} \mathcal{L}_{adv}(G_{Y_{\downarrow}X}, D_{X}, Y_{\downarrow}, X)$.

In our framework, the two generators are also optimized through an HR discriminator $D_{X_{\uparrow}}$:
\begin{equation}
\begin{split}
& \mathcal{L}_{adv}((G_{XY_{\downarrow}}, G_{Y_{\downarrow}X}), D_{X_{\uparrow}}, Y_{\downarrow}, X_{\uparrow})\\
=~& \mathbb{E}_{x \sim {\rm P}_{X}}[\log D_{X_{\uparrow}}(U_{Y_{\downarrow}Y} \circ G_{XY_{\downarrow}}(x))]\\
+~& \mathbb{E}_{y_{\downarrow} \sim {\rm P}_{Y_{\downarrow}}}[\log (1 - D_{X_{\uparrow}}(U_{Y_{\downarrow}Y}(\accentset{\scriptstyle\circ}{y_{\downarrow}})))].
\end{split}
\label{eq:loss_adv_hr}
\end{equation}
The optimization process of Eq.~\ref{eq:loss_adv_hr} is expressed as $\min_{G_{XY_{\downarrow}}, G_{Y_{\downarrow}X}} \max_{D_{X_{\uparrow}}} \mathcal{L}_{adv}((G_{XY_{\downarrow}}, G_{Y_{\downarrow}X}), D_{X_{\uparrow}}, Y_{\downarrow}, X_{\uparrow})$.

\noindent{{\bf Cycle consistency loss.\ }}
The normal CycleGAN learns one-to-one mappings because it imposes cycle consistency on both cycles (\ie, $X \rightarrow Y \rightarrow X$ and $Y \rightarrow X \rightarrow Y$).
We relax this restriction by requiring cycle consistency for only one side:
\begin{equation}
\mathcal{L}_{cyc}(G_{Y_{\downarrow}X}, G_{XY_{\downarrow}}) = \|G_{XY_{\downarrow}} \circ G_{Y_{\downarrow}X}(y_{\downarrow}) - y_{\downarrow}\| _{1}.
\label{eq:loss_cycle}
\end{equation}
Under the above one-side cycle consistency, the mapping $G_{Y_{\downarrow}X}$ is allowed to be one-to-many.
Consequently, our framework can deal with various noise types/distributions of the LR source domain $X$.

\noindent{{\bf Identity mapping loss.\ }}
An identity mapping loss was introduced in the original CycleGAN to preserve color composition for a task of painting~$\rightarrow$~photo.
We also impose the identity mapping loss for $G_{XY_{\downarrow}}$ to avoid color variation:
\begin{equation}
\mathcal{L}_{idt}(G_{XY_{\downarrow}}) = \|G_{XY_{\downarrow}}(y_{\downarrow}) - y_{\downarrow}\| _{1}.
\label{eq:loss_idt}
\end{equation}

\noindent{{\bf Geometric ensemble loss.\ }}
Geometric consistency, which was introduced in a recent work~\cite{fu2019geometry}, reduces the space of possible translation to preserve the scene geometry.
Inspired by geometric consistency, we introduce a simple geometric ensemble loss that requires the flip and rotation for the input images not to change the result:
\begin{equation}
\mathcal{L}_{geo}(G_{XY_{\downarrow}}) = \|G_{XY_{\downarrow}}(x) - \sum_{i=1}^8 T_{i}^{-1}(G_{XY_{\downarrow}}(T_{i}(x)))/8\| _{1},
\label{eq:loss_slf}
\end{equation}
where the operators $\{T_{i}\}_{i=1}^8$ represent eight distinct patterns of flip and rotation.
Note that using $\mathcal{L}_{geo}$ increases the total training time by a factor of approximately $3/2$.

\noindent{{\bf Full objective.\ }}
Our full objective for the two generators and three discriminators is as follows:
\begin{equation}
\begin{split}
\mathcal{L}_{trans}
& =~\mathcal{L}_{adv}(G_{XY_{\downarrow}}, D_{Y_{\downarrow}}, X, Y_{\downarrow})\\
& +~\mathcal{L}_{adv}(G_{Y_{\downarrow}X}, D_{X}, Y_{\downarrow}, X)\\
& +~\gamma \mathcal{L}_{adv}((G_{XY_{\downarrow}}, G_{Y_{\downarrow}X}), D_{X_{\uparrow}}, Y_{\downarrow}, X_{\uparrow})\\
& +~\lambda_{cyc}\mathcal{L}_{cyc}(G_{Y_{\downarrow}X}, G_{XY_{\downarrow}})\\
& +~\lambda_{idt}\mathcal{L}_{idt}(G_{XY_{\downarrow}})\\
& +~\lambda_{geo}\mathcal{L}_{geo}(G_{XY_{\downarrow}}),
\end{split}
\label{eq:loss_full}
\end{equation}
where the hyperparameters $\lambda_{cyc}$, $\lambda_{idt}$, $\lambda_{geo}$, and $\gamma$ weight the contributions of each objective.

While the SR network $U_{Y_{\downarrow}Y}$ is independent of the generators and discriminators, it is used as an amplifier of the local features of images to be inputted to $D_{X_{\uparrow}}$.
Thus, we jointly update the SR network during the training of the correction network.
We use L1 loss to reconstruct an HR image from a {\it pseudo-clean}~LR image $\accentset{\scriptstyle\circ}{y_{\downarrow}}$:
\begin{equation}
\mathcal{L}_{rec} = \|U_{Y_{\downarrow}Y}(\accentset{\scriptstyle\circ}{y_{\downarrow}}) - y\| _{1}.
\label{eq:loss_sr}
\end{equation}
We again note that any pixel-wise loss (\eg, perceptual loss, texture loss, and adversarial loss) can be used as $\mathcal{L}_{rec}$ in our formulation.

\begin{figure}[t]
\centering
\includegraphics[width=0.9\linewidth]{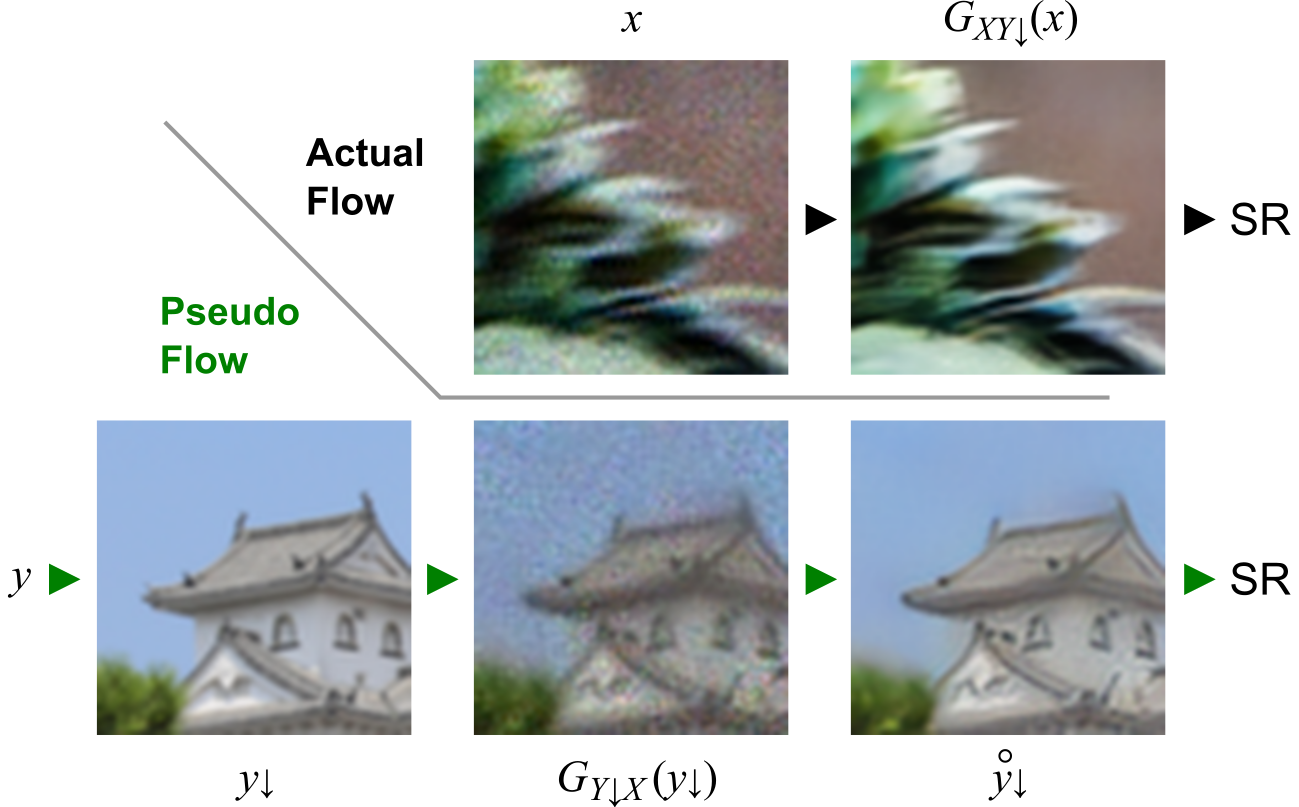}
\caption{{\bf Intermediate images of proposed method.} $x$ is image ``0886'' from the DIV2K realistic-wild validation set, and $y$ is image ``0053'' from the DIV2K training ground-truth set.}
\label{fig:exp1_flow}
\end{figure}

\begin{table*}[t]
\centering
\footnotesize
\begin{tabular}{p{4.0cm}p{9.0cm}p{1.0cm}p{1.0cm}}
\hline
& \hfil \leftline{Method} & \hfil PSNR & \hfil SSIM\\
\hline
\hline
& \hfil \leftline{Bicubic (for reference)} & \hfil 19.99 & \hfil 0.4857\\
\hfil Blind denoising/deblurring & \hfil \leftline{NC~\cite{lebrun2015noise} $+$ Bicubic} & \hfil 20.03 & \hfil 0.5049\\
\hfil $+$ & \hfil \leftline{RL-restore~\cite{yu2018crafting} $+$ Bicubic} & \hfil 20.18 & \hfil 0.5119\\
\hfil Bicubic upscaling & \hfil \leftline{RL-restore~\cite{yu2018crafting} $+$ SRN-Deblur~\cite{tao2018scale} $+$ Bicubic} & \hfil 20.13 & \hfil 0.5173\\
& \hfil \leftline{RL-restore~\cite{yu2018crafting} $+$ DeblurGAN-v2~\cite{kupyn2019deblurgan} $+$ Bicubic} & \hfil 20.21 & \hfil 0.5158\\
\hline
\hfil Blind denoising/deblurring & \hfil \leftline{DBPN \cite{haris2018deep} (for reference)} & \hfil 19.82 & \hfil 0.4572\\
\hfil $+$ non Blind SR method & \hfil \leftline{RL-restore~\cite{yu2018crafting} $+$ DeblurGAN-v2~\cite{kupyn2019deblurgan} $+$ DBPN \cite{haris2018deep}} & \hfil 20.25 & \hfil 0.5198\\
\hline
& \hfil \leftline{ZSSR \cite{shocher2018zero}} & \hfil 19.91 & \hfil 0.4835\\
& \hfil \leftline{ZSSR \cite{shocher2018zero} w/ KernelGAN~\cite{bell2019blind}} & \hfil 19.45 & \hfil 0.4493\\
\hfil Blind denoising/deblurring & \hfil \leftline{IKC \cite{gu2019blind}} & \hfil 19.62 & \hfil 0.4251\\
\hfil $+$ Blind SR method & \hfil \leftline{RL-restore~\cite{yu2018crafting} $+$ DeblurGAN-v2~\cite{kupyn2019deblurgan} $+$ ZSSR \cite{shocher2018zero}} & \hfil 20.19 & \hfil \color{blue}0.5217\\
& \hfil \leftline{RL-restore~\cite{yu2018crafting} $+$ DeblurGAN-v2~\cite{kupyn2019deblurgan} $+$ ZSSR \cite{shocher2018zero} w/ KernelGAN~\cite{bell2019blind}} & \hfil 19.83 & \hfil 0.5137\\
& \hfil \leftline{RL-restore~\cite{yu2018crafting} $+$ DeblurGAN-v2~\cite{kupyn2019deblurgan} $+$ IKC \cite{gu2019blind}} & \hfil \color{blue}20.26 & \hfil 0.5140\\
\hline
& \hfil \leftline{Our method} & \hfil \color{red}21.32 & \hfil \color{red}0.5541\\
\hline
\end{tabular}
\caption{{\bf Numerical comparison with state-of-the-art blind methods on DIV2K realistic-wild validation set (SR scale ${\bf \times 4}$).} The best and second-best results are highlighted in {\color{red} red} and {\color{blue} blue}, respectively. We use the officially provided evaluation script\protect\footnotemark (validation stage setting). Throughout this paper, the real configuration is used for ZSSR, and the Inception backbone model is used for DeblurGAN-v2.}
\label{tab:div2k_sota}
\vspace{-2mm}
\end{table*}

\begin{figure*}[h]
\begin{tabular}{c}
    \begin{minipage}{0.188\hsize}
    \centering
    \includegraphics[width=1.0\linewidth]{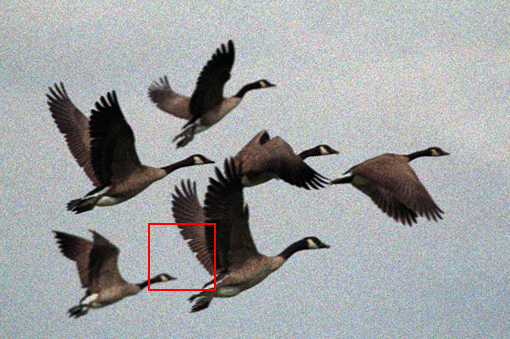}
    \end{minipage}
    
    \begin{minipage}{0.125\hsize}
    \centering
    \includegraphics[width=1.0\linewidth]{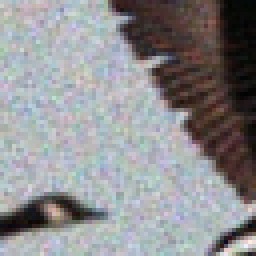}
    \end{minipage}
    
    \begin{minipage}{0.125\hsize}
    \centering
    \includegraphics[width=1.0\linewidth]{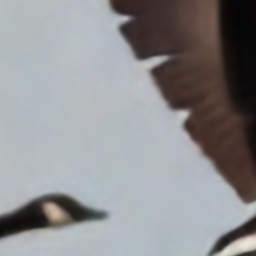}
    \end{minipage}
    
    \begin{minipage}{0.125\hsize}
    \centering
    \includegraphics[width=1.0\linewidth]{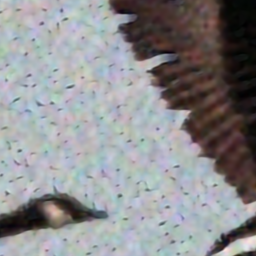}
    \end{minipage}
    
    \begin{minipage}{0.125\hsize}
    \centering
    \includegraphics[width=1.0\linewidth]{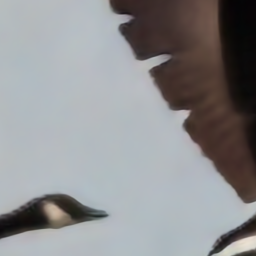}
    \end{minipage}
    
    \begin{minipage}{0.125\hsize}
    \centering
    \includegraphics[width=1.0\linewidth]{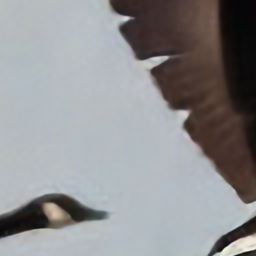}
    \end{minipage}
    
    \begin{minipage}{0.125\hsize}
    \centering
    \includegraphics[width=1.0\linewidth]{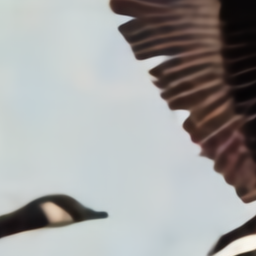}
    \end{minipage}
    \vspace{0.25mm}
\end{tabular}

\begin{tabular}{c}
    \begin{minipage}{0.188\hsize}
    \centering
    \includegraphics[width=1.0\linewidth]{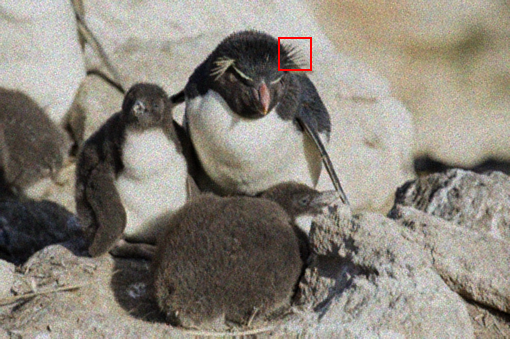}
    \subcaption*{{\footnotesize 
    ``0896'' and ``0842'' from \vspace{-0.75mm}\\
    DIV2K realistic-wild \vspace{-0.75mm}\\
    validation set
    }}
    \end{minipage}
    
    \begin{minipage}{0.125\hsize}
    \centering
    \includegraphics[width=1.0\linewidth]{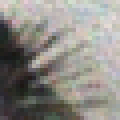}
    \subcaption*{{\footnotesize 
    LR input \vspace{-0.75mm}\\
    ~ \vspace{-0.75mm}\\
    ~
    }\vspace{0.5mm}}
    \end{minipage}
    
    \begin{minipage}{0.125\hsize}
    \centering
    \includegraphics[width=1.0\linewidth]{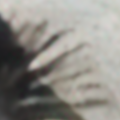}
    \subcaption*{{\footnotesize 
    RL-restore~$+$ \vspace{-0.75mm}\\
    DeblurGAN-v2~$+$ \vspace{-0.75mm}\\
    DBPN
    }}
    \end{minipage}
    
    \begin{minipage}{0.125\hsize}
    \centering
    \includegraphics[width=1.0\linewidth]{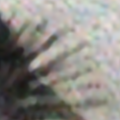}
    \subcaption*{{\footnotesize 
    ZSSR \vspace{-0.75mm}\\
    ~ \vspace{-0.75mm}\\
    ~
    }\vspace{0.5mm}}
    \end{minipage}
    
    \begin{minipage}{0.125\hsize}
    \centering
    \includegraphics[width=1.0\linewidth]{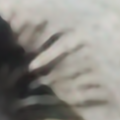}
    \subcaption*{{\footnotesize 
    RL-restore~$+$ \vspace{-0.75mm}\\
    DeblurGAN-v2~$+$ \vspace{-0.75mm}\\
    ZSSR
    }}
    \end{minipage}
    
    \begin{minipage}{0.125\hsize}
    \centering
    \includegraphics[width=1.0\linewidth]{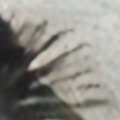}
    \subcaption*{{\footnotesize 
    RL-restore~$+$ \vspace{-0.75mm}\\
    DeblurGAN-v2~$+$ \vspace{-0.75mm}\\
    IKC
    }}
    \end{minipage}
    
    \begin{minipage}{0.125\hsize}
    \centering
    \includegraphics[width=1.0\linewidth]{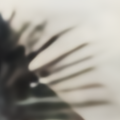}
    \subcaption*{{\footnotesize 
    Ours\vspace{-0.75mm}\\
    ~\vspace{-0.75mm}\\
    ~
    }\vspace{0.5mm}}
    \end{minipage}
\end{tabular}
\caption{{\bf Qualitative comparison with state-of-the-art blind methods on DIV2K realistic-wild validation set (SR scale ${\bf \times 4}$).} Our method reconstructs the fine details while removing artifacts, yielding the most visually pleasing results.}
\label{fig:div2k_compare_blind}
\vspace{-2.5mm}
\end{figure*}

\subsection{Network Architecture}
\noindent{{\bf $\bm{G_{XY_{\downarrow}}}$ and $\bm{U_{Y_{\downarrow}Y}}$.\ }}
We utilize an RCAN~\cite{zhang2018image}-based architecture as $G_{XY_{\downarrow}}$ and $U_{Y_{\downarrow}Y}$.
The RCAN is a very deep SR network realized by a residual in residual structure with short and long skip connections.
The RCAN consists of 10 residual groups (RGs), where each RG contains 20 residual channel attention blocks (RCABs).
Our $G_{XY_{\downarrow}}$ ($U_{Y_{\downarrow}Y}$) is a reduced version of the RCAN consisting of five RGs with 10 (20) RCABs.
Note that the final upscaling layer included in the original RCAN is omitted for $G_{XY_{\downarrow}}$.

\noindent{{\bf $\bm{G_{Y_{\downarrow}X}}$.\ }}
For the generator $G_{Y_{\downarrow}X}$, we use several residual blocks with $5\times5$ filters and several fusion layers with $1\times1$ filters, where each convolution layer is followed by batch normalization (BN)~\cite{ioffe2015batch} and LeakyReLU.
The two head modules, including the one residual block, independently extract the features of an inputted RGB image and single-channel random noise $\mathcal{N}(0, 1)$ that simulates the randomness of distortions.
Then, the two extracted features are concatenated to be inputted to a main module consisting of six residual blocks and three fusion layers.

\noindent{{\bf $\bm{D_X}$, $\bm{D_{Y_{\downarrow}}}$ and $\bm{D_{X_{\uparrow}}}$.\ }}
For the LR discriminators $D_X$ and $D_{Y_{\downarrow}}$, we use five convolution layers with strides of 1.
The convolution layers, except for the last layer, are followed by LeakyReLU without BN.
A similar architecture is also used for the HR discriminator $D_{X_{\uparrow}}$ but with a different stride in the initial layers.
For the case of ${\rm scale~facter} = 2$ ($4$), strides of 2 are used for the first (and second) layer(s) of $D_{X_{\uparrow}}$.
We use PatchGAN~\cite{li2016precomputed, isola2017image} for all the discriminators.

%------------------------------------------------------------------------
\section{Experiments}
%We conducted experiments on four diverse datasets to evaluate the proposed method.

\subsection{Network Training}
We used the Adam optimizer~\cite{kingma2014adam} with $\beta_1 = 0.5$, $\beta_2 = 0.999$, and $\epsilon = 10^{-8}$ to train the generators and discriminators.
The SR network $U_{Y_{\downarrow}Y}$ was similarly trained but with a different $\beta_1$ ($= 0.9$).
The learning rates of all networks were initialized to $1 \times 10^{-4}$.
Then, the learning rates of the networks other than $U_{Y_{\downarrow}Y}$ were halved at 100k, 180k, 240k, and 280k iterations.
We trained our networks for more than $3 \times 10^{5}$ iterations with a mini-batch size of 16.
In each iteration, LR patches of $32\times32$ and HR patches of corresponding size were extracted as inputs in an unaligned manner.
Then, data augmentation of random flip and rotation was performed on each training patch.
We used PyTorch~\cite{paszke2017automatic} to conduct all the experiments.

\subsection{Experiments on Synthetic Distortions}
\label{exp1}
\noindent{{\bf DIV2K realistic-wild dataset.\ }}
We used the realistic-wild set (Track 4) of the NTIRE 2018 Super-Resolution Challenge~\cite{timofte2018ntire}.
The realistic-wild set was generated by degrading DIV2K~\cite{timofte2017ntire}, consists of 2K resolution images that are diverse in their content.
DIV2K has 800 training images.
The realistic-wild set simulates real ``wild'' LR images via $\times$4 downscaling, motion blurring, pixel shifting, and noise addition.
The degradation operations are the same within a single image but vary from image to image.
Four degraded LR images are generated for each DIV2K training image (\ie, 3,200 LR training images in total).
We trained our model using the above 3,200 LR and 800 HR paired images but with ``unpaired/unaligned'' sampling.
We evaluated the results on the 100 realistic-wild validation images because the ground truths of the testing images were not provided.

\noindent{{\bf Hyperparameters.\ }}
We used the loss hyperparameters $\lambda_{cyc} = 1$, $\lambda_{idt} = 1$, $\lambda_{geo} = 1$, and $\gamma = 0.1$ throughout the experiments in this subsection.
The SR factor was $\times 4$.

\noindent{{\bf Intermediate images.\ }}
Figure \ref{fig:exp1_flow} shows visual examples of the intermediate images of the proposed method.
The degradation network $G_{Y_{\downarrow}X}$ degrades the input {\it clean}~LR image $y_{\downarrow}$ so that the output $G_{Y_{\downarrow}X}(y_{\downarrow})$ reproduces the noise distribution of the real degraded image $x$.
The reconstruction network $G_{XY_{\downarrow}}$ is effective in removing the noises of both the real ($x$) and fake ($G_{Y_{\downarrow}X}(y_{\downarrow})$) degraded images.

\footnotetext[2]{\scriptsize {\url{https://data.vision.ee.ethz.ch/cvl/DIV2K/}}}
\noindent{{\bf Comparison with state-of-the-art blind methods.\ }}
Because the blind SR method for multiple degradations has not been studied sufficiently, we took a benchmark by combining the SR method with the blind restoration methods (Tab.~\ref{tab:div2k_sota}, Fig.~\ref{fig:div2k_compare_blind}).
We first explored the state-of-the-art blind denoising methods: a patch-based method NC~\cite{lebrun2015noise} and a CNN-based method RL-restore~\cite{yu2018crafting}.
RL-restore performed better than NC.
Then, we compared two CNN-based blind deblurring methods, SRN-Deblur~\cite{tao2018scale} and DeblurGAN-v2~\cite{kupyn2019deblurgan}, based on the output of RL-restore.
The performances of these deblurring methods were almost equivalent, but DeblurGAN-v2 ran faster.
Finally, three state-of-the-art SR methods were combined with RL-restore and DeblurGAN-v2: a non-blind SR method DBPN~\cite{haris2018deep} and two blind SR methods ZSSR~\cite{shocher2018zero} and IKC~\cite{gu2019blind}.
We further combined ZSSR with the recently proposed kernel estimation method KernelGAN~\cite{bell2019blind}.
Our method outperformed all of the above methods by a large margin; however, the comparison was not completely fair because the compared methods were not trained on the dataset used here.

\begin{table}
\centering
\footnotesize
\begin{tabular}{p{3.2cm}p{0.8cm}p{0.8cm}}
\hline
\hfil \leftline{Username} & \hfil PSNR & \hfil SSIM\\
\hline
\hline
\hfil \leftline{xixihaha} & \hfil \color{red}24.12 & \hfil \color{red}0.56\\
\hfil \leftline{yyuan13} & \hfil \color{blue}24.07 & \hfil \color{red}0.56\\
\hfil \leftline{Hot\_Milk} & \hfil 23.90 & \hfil \color{red}0.56\\
\hfil \leftline{yifita} & \hfil 23.87 & \hfil \color{red}0.56\\
\hfil \leftline{cskzh} & \hfil 23.55 & \hfil \color{blue}0.55\\
\hfil \leftline{JSChoi} & \hfil 23.20 & \hfil 0.53\\
\hfil \leftline{enoch} & \hfil 23.04 & \hfil 0.52\\
\hfil \leftline{assafsho} & \hfil 22.93 & \hfil 0.51\\
\hfil \leftline{hyu\_ss} & \hfil 22.57 & \hfil 0.49\\
\hfil \leftline{cr2018} & \hfil 22.52 & \hfil 0.49\\
\hline
\hfil \leftline{Ours} & \hfil 21.32 & \hfil \color{blue}0.5541\\
\hfil \leftline{Ours$^+$} & \hfil 21.35 & \hfil \color{red}0.5560\\
\hline
\end{tabular}
\caption{{\bf Comparison with NTIRE 2018 baselines.} Top 10 validation results from NTIRE 2018 realistic-wild challenge website are compared as pair-trained upper bounds. Ours$^+$ is an enhanced version of Ours using a standard self-ensemble technique~\cite{timofte2016seven}.}
\label{tab:div2k_ntire}
\vspace{-2mm}
\end{table}

\begin{table}
\centering
\footnotesize
\begin{tabular}{p{5.2cm}p{0.8cm}p{0.8cm}}
\hline
\hfil \leftline{Method} & \hfil PSNR & \hfil SSIM\\
\hline
\hline
\hfil \leftline{Ours} & \hfil 21.32 & \hfil 0.5541\\
\hfil \leftline{Ours - w/o $D_{X_{\uparrow}}$} & \hfil 21.29 & \hfil 0.5532\\
\hfil \leftline{Ours - trained on $y_{\downarrow}$} & \hfil 21.09 & \hfil 0.5312\\
\hfil \leftline{Ours - trained on $G_{Y_{\downarrow}X}(y_{\downarrow})$} & \hfil 20.84 & \hfil 0.5500\\
\hfil \leftline{Ours - trained on $G_{Y_{\downarrow}X}(y_{\downarrow})$ - original RCAN} & \hfil 20.78 & \hfil 0.5482\\
\hline
\end{tabular}
\caption{{\bf Ablation study.} Some other variants of our network were compared to verify the proposed method.}
\label{tab:div2k_ablation}
\end{table}

\noindent{{\bf Comparison with NTIRE 2018 baselines.\ }}
Table \ref{tab:div2k_ntire} shows a comparison with NTIRE 2018 baselines from the validation website\footnote{\scriptsize {\url{https://competitions.codalab.org/competitions/18026}}}, where the dataset and evaluation script used were the same as in our experiment.
Note that although the NTIRE 2018 competition provides a paired training dataset, we trained our network in an unpaired manner.
Thus, the NTIRE 2018 baselines can be regarded as pair-trained upper bounds.
Our result is inferior to the upper bounds in PSNR, but the result of the more sophisticated indicator SSIM~\cite{wang2004image} is comparable to the upper bounds.
Because PSNR overestimates slight differences in global brightness and/or color that do not significantly affect the perceptual quality~\cite{wang2009mean}, we believe our method shows practically equivalent performance to the pair-trained upper bounds.

\noindent{{\bf Ablation study.\ }}
To investigate the effectiveness of the proposed method, we designed some other variants of our network: (1) Ours - w/o $D_{X_{\uparrow}}$, where the HR discriminator $D_{X_{\uparrow}}$ is removed (\ie $\gamma = 0$), (2) Ours - trained on $y_{\downarrow}$, where the SR network $U_{Y_{\downarrow}Y}$ is trained on $y_{\downarrow}$ instead of $\accentset{\scriptstyle\circ}{y_{\downarrow}}$, which is equivalent to a simple combination of a style translation network and SR network, and (3) Ours - trained on $G_{Y_{\downarrow}X}(y_{\downarrow})$, where the SR network $U_{Y_{\downarrow}Y}$ is trained on $G_{Y_{\downarrow}X}(y_{\downarrow})$ instead of $\accentset{\scriptstyle\circ}{y_{\downarrow}}$ and only $U_{Y_{\downarrow}Y}$ is used at testing time, which is equivalent to the indirect approach illustrated in Fig.~\ref{fig:two_approaches_b}.
For completeness, the variant (3) was validated using the original RCAN model as the SR network, which is larger than our total testing network $U_{Y_{\downarrow}Y} \circ G_{XY_{\downarrow}}$.
These variants underperformed compared to the proposed method (Tab.~\ref{tab:div2k_ablation}).
In particular, our full model outperformed variant (3), meaning that the proposed {\it pseudo-supervision} is effective at reducing the train--test discrepancy.

\begin{figure}[t]
\centering
\includegraphics[width=0.9125\linewidth]{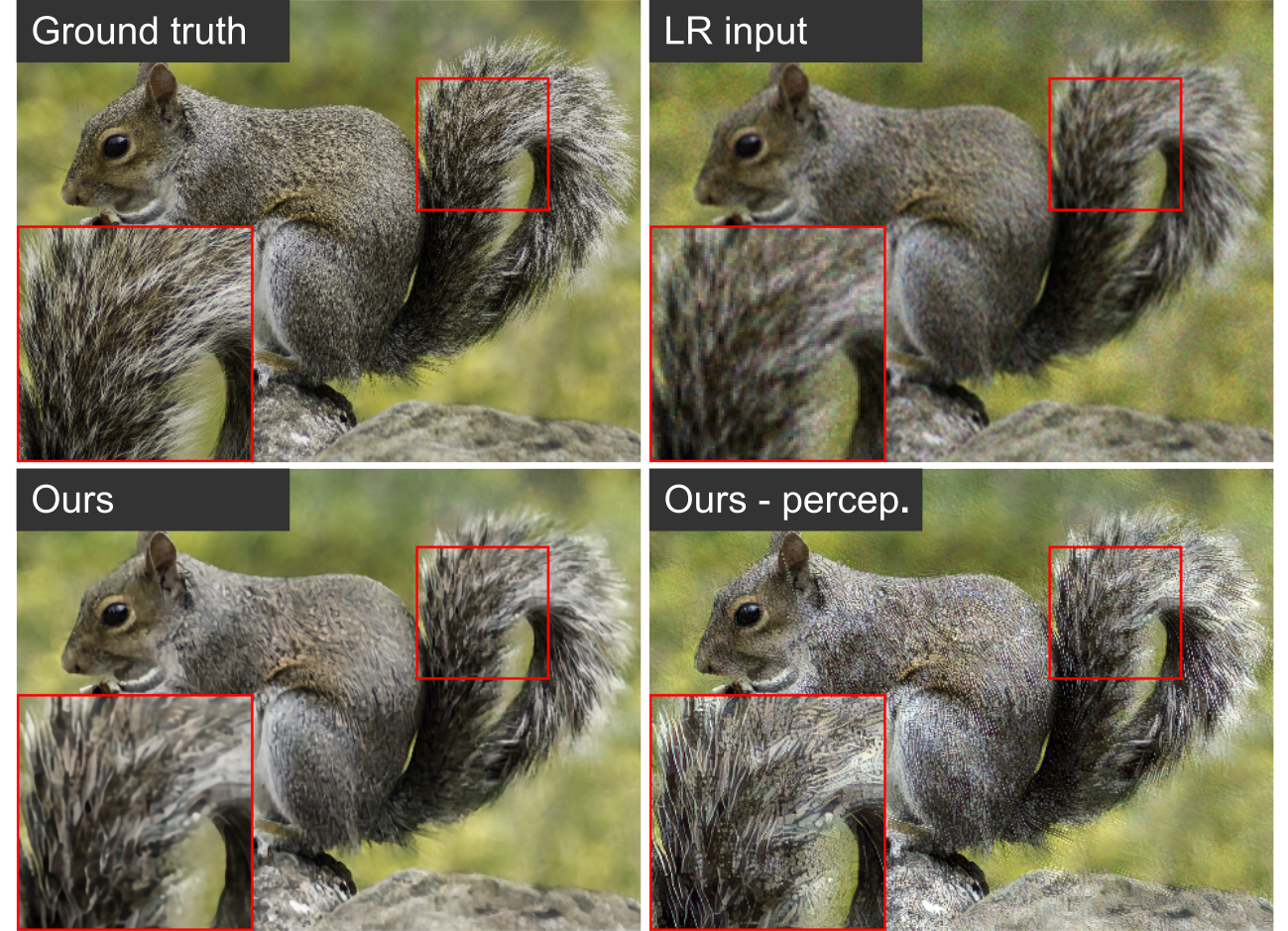}
\caption{{\bf Example image with perception-oriented training.} ``0810'' from DIV2K realistic-wild validation set is displayed.}
\label{fig:perceptual}
\end{figure}

\noindent{{\bf Perception-oriented training.\ }}
We also trained our model with a perception-oriented reconstruction loss following ESRGAN~\cite{wang2018esrgan} to demonstrate the versatility of our method.
We replaced Eq.~\ref{eq:loss_sr} with a combination of perceptual loss, relativistic adversarial loss~\cite{jolicoeur2018relativistic}, and content loss as in ESRGAN, while the other loss functions and training procedure were unchanged.
The perceptually trained model gives a more visually pleasing result than the normal model trained with L1 reconstruction loss (Fig.~\ref{fig:perceptual}).

\subsection{Experiments on Realistic Distortions I}
\label{exp2}
\noindent{{\bf Large-scale face image dataset.\ }}
In this subsection, we follow the experimental procedure described by Bulat~\etal~\cite{bulat2018learn}.
We used the dataset and evaluation script they provided\footnote{\scriptsize {\url{https://github.com/jingyang2017/Face-and-Image-super-resolution}}}.
They collected 182,866 HR face images from the Celeb-A~\cite{liu2015deep}, AFLW~\cite{koestinger2011annotated}, LS3D-W~\cite{bulat2017far}, and VGGFace2~\cite{cao2018vggface2}.
They also collected more than 50,000 real-world LR face images from Widerface~\cite{yang2016wider} that are diverse in degradation types.
3,000 images were randomly selected from the LR dataset and kept for testing.
Then, all HR and LR face images were cropped in a consistent manner using the face detector~\cite{zhang2017s3fd}.
The cropped HR and LR training images were $64 \times 64$ and $16 \times 16$ patches, respectively.

\begin{figure}[t]
\centering
\includegraphics[width=0.975\linewidth]{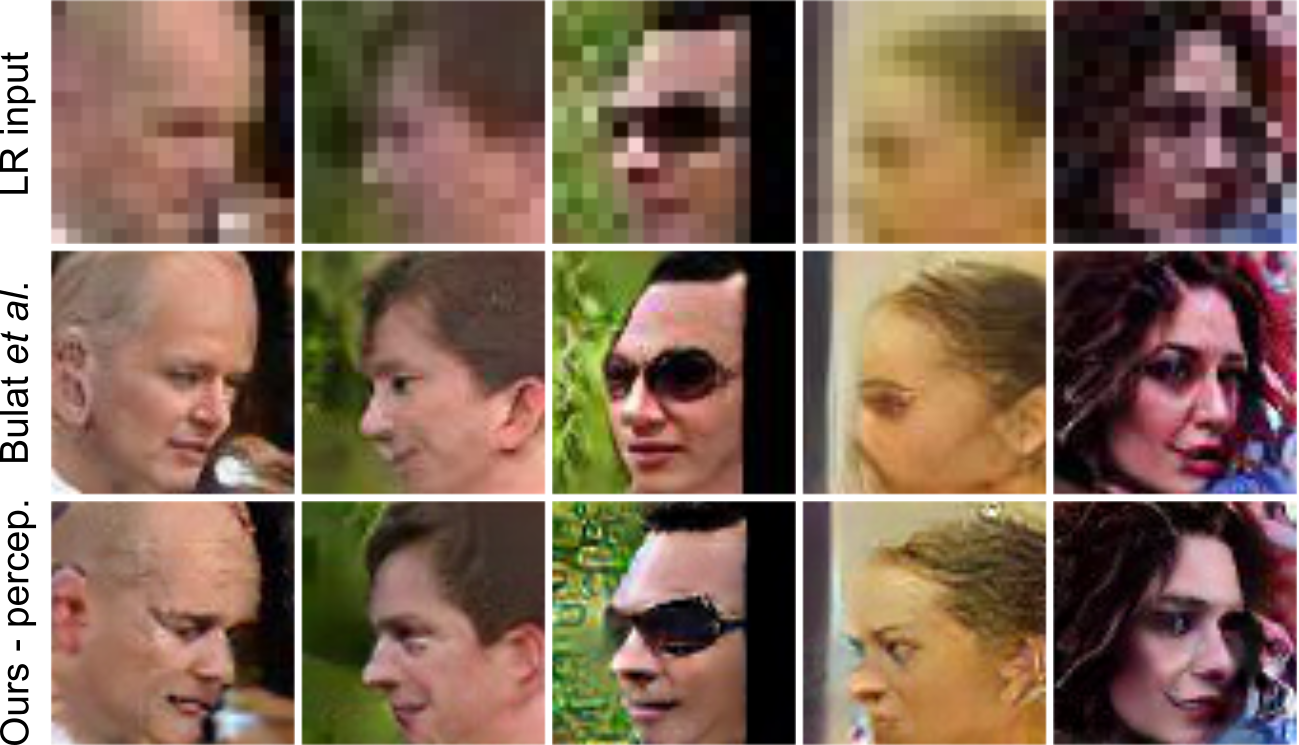}
\caption{{\bf Qualitative comparison with a state-of-the-art GAN-based unpaired SR method (SR scale ${\bf \times 4}$).} Input LR images are from the LR test set provided by Bulat~\etal.}
\label{fig:face1}
\vspace{-2mm}
\end{figure}

\begin{table}
\centering
\footnotesize
\begin{tabular}{p{4.8cm}p{0.8cm}}
\hline
\hfil \leftline{Method} & \hfil FID\\
\hline
\hline
\hfil \leftline{SRGAN~\cite{ledig2017photo}} & \hfil 104.80\\
\hfil \leftline{CycleGAN~\cite{zhu2017unpaired}} & \hfil 19.01\\
\hfil \leftline{DeepDeblur~\cite{nah2017deep}} & \hfil 294.96\\
\hfil \leftline{Wavelet-SRNet~\cite{huang2017wavelet}} & \hfil 149.46\\
\hfil \leftline{FSRNet~\cite{chen2018fsrnet}} & \hfil 157.29\\
\hfil \leftline{Bulat~\etal~\cite{bulat2018learn}} & \hfil \color{blue}14.89\\
\hline
\hfil \leftline{Ours - perceptual} & \hfil \color{red}13.57\\
\hline
\end{tabular}
\caption{{\bf FID-based performance comparison with state-of-the-art methods.} The dataset and evaluation script provided by Bulat~\etal were used. Lower scores indicate better results.}
\label{tab:face_sota}
\end{table}

\noindent{{\bf Hyperparameters.\ }}
For the experiments on realistic distortions, we found that it is better to take $x$ instead of $y_{\downarrow}$ as an argument of the identity mapping loss.
Thus, we used the modified identity mapping loss
\begin{equation}
\mathcal{L}_{\overline{idt}}(G_{XY_{\downarrow}}) = \|G_{XY_{\downarrow}}(x) - x\| _{1}
\label{eq:loss_idt_mod}
\end{equation}
instead of Eq.~\ref{eq:loss_idt} in the following.
We used the loss hyperparameters $\lambda_{cyc} = 1$, $\lambda_{\overline{idt}} = 2$, $\lambda_{geo} = 1$, and $\gamma = 0.1$.
We first upscaled all the $16 \times 16$ LR patches by a factor of two using a bicubic method because the original size was too small.
Then, our network was trained on the $32 \times 32$ LR patches and $64 \times 64$ HR patches with an SR factor of $\times 2$.

\noindent{{\bf Comparison with state-of-the-art methods.\ }}
We numerically and qualitatively compared our method with the state-of-the-art GAN-based unpaired method proposed by Bulat~\etal~\cite{bulat2018learn}.
Our method was also numerically compared with five related state-of-the-art methods: image SR method SRGAN~\cite{ledig2017photo}, face SR methods Wavelet-SRNet~\cite{huang2017wavelet} and FSRNet~\cite{chen2018fsrnet}, unpaired image translation method CycleGAN~\cite{zhu2017unpaired}, and deblurring method DeepDeblur~\cite{nah2017deep}.
Please see Ref.~\cite{bulat2018learn} for a more detailed explanation of each method.

Table~\ref{tab:face_sota} shows a numerical comparison with the related state-of-the-art methods.
We assessed the quality of the SR results with the Fr\'echet inception distance (FID)~\cite{heusel2017gans} because there were no corresponding ground-truth images.
CycleGAN, Bulat~\etal's, and our method, which are GAN-based unpaired approaches, largely outperformed all other methods.
Besides, our method showed better performance than CycleGAN and Bulat~\etal's.
For completeness, we calculated PSNR between the bicubically upscaled LR test images and its SR results.
The calculated PSNRs for the results of Bulat~\etal and our method were 20.28 dB and 21.09 dB, respectively.
These numerical results indicate that our method produces perceptually better results than Bulat~\etal's while maintaining the characteristics of the input images.
A qualitative comparison is shown in Fig.~\ref{fig:face1}.

\begin{figure}[t]
\centering
\includegraphics[width=1.0\linewidth]{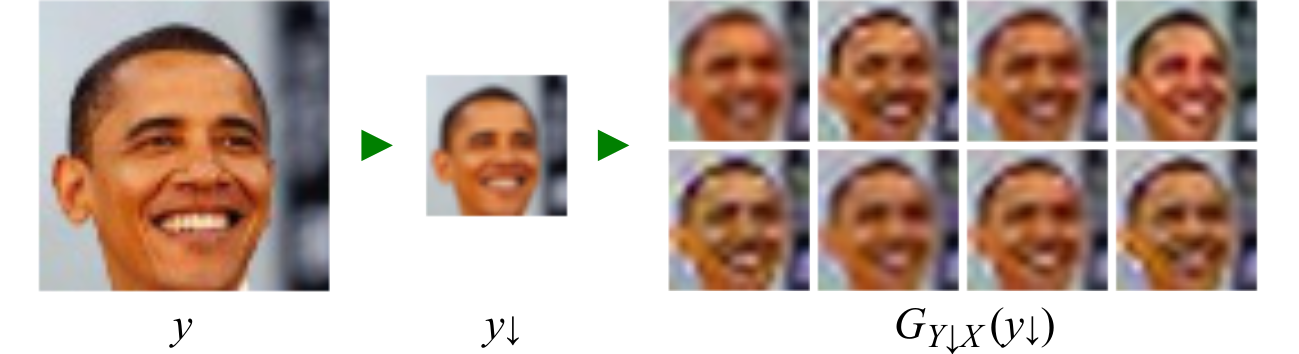}
\caption{{\bf One-to-many degradation examples.} Examples of different LR images generated by our degradation network for different random noise input.}
\label{fig:face2}
\end{figure}

\noindent{{\bf One-to-many degradation examples.\ }}
Visual examples expressing the various noise intensities/types learned by our degradation network $G_{Y_{\downarrow}X}$ shown in Fig.~\ref{fig:face2}.
The one-sided cycle consistency allows the mapping $G_{Y_{\downarrow}X}$ to become one-to-many, reproducing the various noise distributions of the real LR images.

\begin{figure*}[h]
\begin{tabular}{c}
    \begin{minipage}{0.158\hsize}
    \centering
    \includegraphics[width=1.0\linewidth]{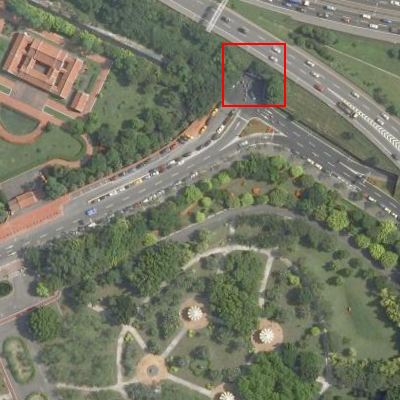}
    \subcaption*{{\footnotesize ``P2331'' from DOTA}}
    \end{minipage}
    
    \begin{minipage}{0.158\hsize}
    \centering
    \includegraphics[width=1.0\linewidth]{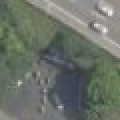}
    \subcaption*{{\footnotesize LR input}}
    \end{minipage}
    
    \begin{minipage}{0.158\hsize}
    \centering
    \includegraphics[width=1.0\linewidth]{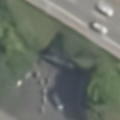}
    \subcaption*{{\footnotesize RL-restore~$+$~Bicubic}}
    \end{minipage}
    
    \begin{minipage}{0.158\hsize}
    \centering
    \includegraphics[width=1.0\linewidth]{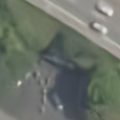}
    \subcaption*{{\footnotesize RL-restore~$+$~DBPN}}
    \end{minipage}
    
    \begin{minipage}{0.158\hsize}
    \centering
    \includegraphics[width=1.0\linewidth]{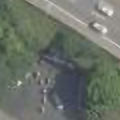}
    \subcaption*{{\footnotesize ZSSR}}
    \end{minipage}
    
    \begin{minipage}{0.158\hsize}
    \centering
    \includegraphics[width=1.0\linewidth]{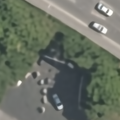}
    \subcaption*{{\footnotesize Ours}}
    \end{minipage}
\end{tabular}
\caption{{\bf Qualitative comparison with state-of-the-art blind methods on DOTA validation set (SR scale ${\bf \times 2}$).} Zoom in for better view.}
\label{fig:dota_compare_blind}
\vspace{-2.5mm}
\end{figure*}

\subsection{Experiments on Realistic Distortions I\hspace{-.1em}I}
\label{exp3}
\noindent{{\bf DOTA and Inria aerial image dataset.\ }}
We used two aerial image datasets with different ground sample distances (GSD) as source and target.
We sampled 62 LR source images with the GSD in the range [55cm, 65cm] from a training set of DOTA~\cite{xia2018dota} (a large-scale aerial image dataset for object detection collected from different sensors and platforms).
For the HR target, we used the Inria aerial image labeling dataset~\cite{maggiori2017can}, which contains scenes from several different cities but with the same GSD (30 cm).
Note that we used only the images of Vienna city (36 images) so that the qualities of the target images are constant.

\noindent{{\bf Hyperparameters.\ }}
We used the loss hyperparameters $\lambda_{cyc} = 1$, $\lambda_{\overline{idt}} = 10$, $\lambda_{geo} = 100$, and $\gamma = 0.1$.
The SR factor was $\times 2$.
%Here, we used larger $\lambda_{\overline{idt}}$ and $\lambda_{geo}$ than in the other experiments.
%This is because, in aerial images, the pixel sizes of objects such as vehicles and buildings are rather small, thus it is necessary to maintain the local characteristics of the images.
In aerial images, the pixel sizes of objects such as vehicles and buildings are rather small, thus we used larger $\lambda_{\overline{idt}}$ and $\lambda_{geo}$ than in the other experiments to maintain the local characteristics of the images.
We gradually elevated $\lambda_{geo}$ in the early stages of training to avoid a mode where the entire image is uniform.

\noindent{{\bf Comparison with state-of-the-art methods.\ }}
We only provide a qualitative comparison in this subsection (Fig.~\ref{fig:dota_compare_blind}) because there are no ground-truth HR images.
The input LR image was sampled from the DOTA validation set with the GSD in the range of [55cm, 65cm].
As the benchmark, a CNN-based blind denoising method RL-restore~\cite{yu2018crafting} was first tested because the input LR images contained visible artifacts.
RL-restore successfully removed the artifacts, but the fine details of the inputted images were removed as well.
The over-smoothed output of RL-restore is slightly enhanced by applying a state-of-the-art SR method DBPN~\cite{haris2018deep}.
A state-of-the-art blind SR method ZSSR~\cite{shocher2018zero} was also tested, but the artifacts were not completely removed.
Unlike the above methods, our method super-resolves the fine details while removing the artifacts, yielding the most visually reasonable results.

\begin{figure}[t]
\begin{tabular}{c}
    \begin{minipage}{0.307\hsize}
    \centering
    \includegraphics[width=1.0\linewidth]{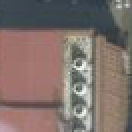}
    \subcaption*{{\footnotesize LR input}}
    \end{minipage}
    
    \begin{minipage}{0.307\hsize}
    \centering
    \includegraphics[width=1.0\linewidth]{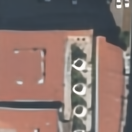}
    \subcaption*{{\footnotesize w/o $\mathcal{L}_{geo}$}}
    \end{minipage}
    
    \begin{minipage}{0.307\hsize}
    \centering
    \includegraphics[width=1.0\linewidth]{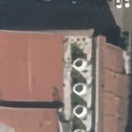}
    \subcaption*{{\footnotesize w/ $\mathcal{L}_{geo}$}}
    \end{minipage}
\end{tabular}
\caption{{\bf Effect of geometric ensemble loss.} ``P2768'' from DOTA validation set is displayed as example.}
\label{fig:dota_geo}
\end{figure}

\noindent{{\bf Effect of geometric ensemble loss.\ }}
We compared the SR results with and without the geometric ensemble loss $\mathcal{L}_{geo}$ to confirm the effectiveness of $\mathcal{L}_{geo}$.
Example images for visual comparison are shown in Fig.~\ref{fig:dota_geo}.
It can be seen that the method without $\mathcal{L}_{geo}$ produces geometrically inconsistent results.
By enforcing geometry consistency through $\mathcal{L}_{geo}$, our method results in more reasonable mapping, preserving geometrical structures of the input LR image.

\subsection{Additional Experiment}
\label{exp4}
We conducted an additional experiment on the dataset provided in the recent AIM 2019 Real-World Super-Resolution Challenge, where no training HR--LR image pairs are available.
We focused on Track 2 of the challenge which is a more general setting than Track 1 (see the competition website\footnote{\scriptsize {\url{https://competitions.codalab.org/competitions/20164}}} for more details).
We compared our method with Lugmayr~\etal~\cite{lugmayr2019unsupervised} which is a study of a GAN-based unpaired SR method (indirect approach; Fig.~\ref{fig:two_approaches_b}) recently published by the organizers of the challenge.
As shown in Tab.~\ref{tab:aim_challenge}, our method achieved superior scores in both the distortion metrics (PSNR, SSIM) and the perception metric (LPIPS~\cite{zhang2018unreasonable}; lower is better).
The visual results are provided in the supplemental material.

\begin{table}[t]
\centering
\footnotesize
\begin{tabular}{p{3.2cm}p{0.8cm}p{0.8cm}p{0.8cm}}
\hline
\hfil \leftline{Method} & \hfil PSNR & \hfil SSIM & \hfil LPIPS\\
\hline
\hline
\hfil \leftline{ZSSR~\cite{shocher2018zero}} & \hfil 22.42 & \hfil 0.61 & \hfil 0.5996\\
\hfil \leftline{ESRGAN~\cite{wang2018esrgan}} & \hfil 20.69 & \hfil 0.51 & \hfil 0.5604\\
\hfil \leftline{Lugmayr~\etal~\cite{lugmayr2019unsupervised}} & \hfil 21.59 & \hfil 0.55 & \hfil 0.4720\\
\hline
\hfil \leftline{Ours} & \hfil \color{blue}22.88 & \hfil \color{blue}0.6612 & \hfil \color{red}0.4539\\
\hfil \leftline{Ours$^+$} & \hfil \color{red}23.01 & \hfil \color{red}0.6655 & \hfil \color{blue}0.4567\\
%\hfil \leftline{Ours - perceptual} & \hfil 19.62 & \hfil 0.5359 & \hfil \color{red}0.3727\\
\hline
\end{tabular}
\caption{{\bf Additional experiment on AIM 2019 Real-World Super-Resolution Challenge dataset (Track 2).} An officially provided evaluation script was used for PSNR and SSIM calculations. A version of the LPIPS script used to evaluate our method is v0.1 (v0.0 outputs lower LPIPS value). We used the loss hyperparameters $\lambda_{cyc} = 1$, $\lambda_{\overline{idt}} = 5$, $\lambda_{geo} = 1$, and $\gamma = 0.1$.}
\label{tab:aim_challenge}
\end{table}

%-------------------------------------------------------------------------
\section{Conclusion}
We investigated the SR problem in an unpaired setting where the aligned HR--LR training set is unavailable.
Our network produces {\it pseudo-clean}~LR images as the intermediate products from ground-truth HR images, which are then used to train the SR network in a paired manner (referred to as ``{\it pseudo-supervision}'' in this paper).
In this sense, the proposed method bridges the gap between the well-studied existing SR methods and the real-world SR problem without paired datasets.
The effectiveness of our method was demonstrated by extensive experiments on diverse datasets: synthetically degraded natural images (Sec.~\ref{exp1}, \ref{exp4}), real-world face images (Sec.~\ref{exp2}), and real-world aerial images (Sec.~\ref{exp3}).

While the proposed method is applicable to diverse datasets, hyperparameter tuning is necessary for each case to maximize the performance.
Making the network more robust against the hyperparameters will be future work.

%-------------------------------------------------------------------------
\section*{Acknowledgement}
I thank Tatsuya Nagata, Shunsuke Ono, Kazuki Sekine, Hiraku Shibuya and Yusuke Uchida for helpful comments on the manuscript.

%-------------------------------------------------------------------------
{\small
\bibliographystyle{ieee_fullname}
\bibliography{ms}
}

\end{document}